\UseRawInputEncoding
\documentclass[runningheads,a4paper]{llncs}

\usepackage{amssymb}
\usepackage{graphicx}

\usepackage{booktabs}
\usepackage{url}
\usepackage{algpseudocode}
\usepackage{amsmath}
\usepackage{amsfonts}
\usepackage[ruled,linesnumbered]{algorithm2e}
\usepackage{multicol}     
\usepackage{multirow}
\usepackage[misc]{ifsym}
\usepackage{color}
\usepackage{cite}
\usepackage{float}
\usepackage{array}

\urldef{\mailsa}\path|{alfred.hofmann, ursula.barth, ingrid.haas, frank.holzwarth,|
\urldef{\mailsb}\path|anna.kramer, leonie.kunz, christine.reiss, nicole.sator,|
\urldef{\mailsc}\path|erika.siebert-cole, peter.strasser, lncs}@springer.com|    
\newcommand{\keywords}[1]{\par\addvspace\baselineskip
\noindent\keywordname\enspace\ignorespaces#1}

\begin{document}

\mainmatter  

\title{Ponzi Scheme Detection in Ethereum Transaction Network}

\titlerunning{Ponzi Scheme Detection in Ethereum Transaction Network}

%
%

\author{Shanqing Yu\textsuperscript{1,2} \and
        Jie Jin\textsuperscript{1,2}     \and
        Yunyi Xie\textsuperscript{1,2}       \and
        Jie Shen\textsuperscript{1,2}    \and
        Qi Xuan\textsuperscript{1,2,3}\textsuperscript{(\Letter)}
}
\institute{       
\textsuperscript{1} Institute of Cyberspace Security, Zhejiang University of Technology, \\Hangzhou 310023, China \\
\textsuperscript{2} College of Information Engineering, Zhejiang University of Technology, \\Hangzhou 310023, China\\
\textsuperscript{3} PCL Research Center of Networks and Communications, Peng Cheng Laboratory, \\Shenzhen 518000, China\\
\email{xuanqi@zjut.edu.cn}
}
\authorrunning{J. Jin et al.}
 
%
%

\toctitle{Lecture Notes in Computer Science}
\tocauthor{Authors' Instructions}
\maketitle

\begin{abstract}
With the rapid growth of blockchain, an increasing number of users have been attracted and many implementations have been refreshed in different fields. Especially in the cryptocurrency investment field, blockchain technology has shown vigorous vitality. However, along with the rise of online business, numerous fraudulent activities, e.g., money laundering, bribery, phishing, and others, emerge as the main threat to trading security. Due to the openness of Ethereum, researchers can easily access Ethereum transaction records and smart contracts, which brings unprecedented opportunities for Ethereum scams detection and analysis. This paper mainly focuses on the Ponzi scheme, a typical fraud, which has caused large property damage to the users in Ethereum. By verifying Ponzi contracts to maintain Ethereum's sustainable development, we model Ponzi scheme identification and detection as a node classification task. In this paper, we first collect target contracts' transactions to establish transaction networks and propose a detecting model based on graph convolutional network~(GCN) to precisely distinguish Ponzi contracts. Experiments on different real-world Ethereum datasets demonstrate that our proposed model has promising results compared with general machine learning methods to detect Ponzi schemes.

\end{abstract}

\section{Introduction}
As the largest decentralized distributed electronic ledger, blockchain has the characteristics of decentralization, non-tampering, transparency, and traceability. Based on the above characteristics, blockchain not only lays the foundation for its trust but also creates a reliable cooperation mechanism for transactions. In a nutshell, blockchain achieves peer-to-peer trust~\cite{swan2015blockchain} without an intermediate third party's credit endorsement, i.e., the transaction is supervised by the whole blockchain, and transaction records cannot misrepresent. There is no doubt that blockchain has broad application prospects. 

Ethereum is an open-source public blockchain with smart contracts, which is called blockchain 2.0~\cite{buterin2014next,wood2014ethereum}. Comparing with Bitcoin, Ethereum not only supports smart contracts but also allows to create and utilize decentralized applications~\cite{wu2019empirical}. With the development of blockchain technology, blockchain has evolved into two circles, i.e., ``chain circle" and ``coin circle". The ``chain circle" focuses on the development and application of blockchain technology while the ``coin circle" is related to digital cryptocurrency. Along with the flourishing of ``coin circle" in digital cryptocurrency market, various fraudulent activities have been arisen, among which the most representative is Ponzi scheme. The traditional Ponzi scheme is that old investors get capital returns from new investors until the investment project is unsustainable, leading to the collapse of project~\cite{frankel2012ponzi}. Comparing with traditional scenarios, the combination of Ponzi scheme and blockchain technology in Ethereum has produced more fresh forms~\cite{vasek2015there}, implement fraud via deployment-and-execution smart contracts~\cite{juels2016ring}. According to a report of~\textit{Cryptoanalysis}, a provider of investigation and risk analysis company for virtual currency, at least $725$ million has been lost in frauds such as Ponzi scheme, ICO revocation and fraud ICO. Until now, cryptocurrencies such as Bitcoin are facing a new round of high value, attracting many inexperienced investors into the ``coin circle", but most have been suffering from a variety of scams. In summary, it's indicated that financial security has become a critical issue in the blockchain ecosystem. Identifying Ponzi scheme in Ethereum can not only protect the interests of investors promptly and reduce the losses of investors but also strengthen the supervision of fraudulent activities in the ``coin circle". Therefore, we mainly focus on Ponzi schemes, which has a very negative impact on Ethereum. 

At present, many scholars have done relevant research on Ponzi scheme detection~\cite{bartoletti2018data}. Most of them are based on machine learning methods which are nothing more than manually extracting contract transaction features, calculating contract code similarity, and counting the frequency of contract opcodes~\cite{chen2018detecting,chen2019exploiting}. However, these methods are mostly based on feature engineering, but pay little attention to the topological structure of transaction networks, which may lose lots of details. In this paper, the problem of identifying Ponzi contracts in Ethereum transaction networks is modeled as a node classification task. Specifically, we propose a comprehensive identification model for the detection of Ponzi schemes in Ethereum. Due to the openness of Ethereum, we can easily obtain all the transaction records of the target smart contracts to establish transaction networks. The external accounts and smart contracts are regarded as nodes, the edges present the transaction between the two, and the node features are extracted from the transactions to facilitate the design of the following algorithm. Distinct from the traditional feature extraction method considering network topology, we apply the graph convolutional network~(GCN) combined structural information with node features to solve an urgent but less studied security issue, i.e., Ponzi scheme in Ethereum.

The main contributions of this paper are summarized as follows. 
\begin{itemize}
\item[$\bullet$] We study the matter of Ponzi scheme detection in Ethereum from a network perspective, which is crucial for Ethereum's sustainable development. In addition, the data used in this paper has been cleaned up and standardized, it will be available online for future study.
\item[$\bullet$] We manually extract inherent characteristics of Ponzi smart contracts, and then incorporate these features with topological structure of transaction networks to design our Ponzi scheme detection method based on GCN.
\item[$\bullet$] Extensive experiments on real-world Ethereum datasets validate the effectiveness of our proposed model on identifying Ponzi schemes, compared with a series of machine learning methods.
\end{itemize}

The rest of the paper is organized as follows. In Section~\ref{sec:RelatedWork}, we review the related works of Ethereum, the detection of Ponzi scheme, and some network embedding methods. In Section~\ref{sec:Methodology}, we present the data collection, contracts' features, and our proposed   Ponzi scheme detection model. In Section~\ref{sec:Experiment}, we conduct extensive experiments with discussions. In Section~\ref{sec:Conclusion}, we conclude the paper and highlight future work.

\section{Related Work} \label{sec:RelatedWork}
\subsection{Ponzi Schemes on Ethereum}
Ethereum~\cite{buterin2014next} that supports smart contracts is currently considered as the second-generation blockchain, and the corresponding cryptocurrency Ether~(ETH) is the second-largest cryptocurrency. However, due to the high open-source nature of Ethereum and the convenience of using smart contracts, many fraudulent activities breed along with Ethereum's high-speed development. Initial coin offering~(ICO) is the most classic application of smart contracts for the blockchain industry, which refers to financing through the issuance of tokens~\cite{hahn2018initial}. Some projects with practical significance can get project start-up funds quickly through ICO financing, but 10\% of whole Ethereum smart contracts are Ponzi schemes and it's difficult for inexperienced investors to distinguish. In order to purify the investment environment of the cryptocurrency platform and reduce the investment loss of investors~\cite{morris2017rise, kondor2014rich}, many researchers have launched a lot of meaningful works to detect Ponzi scheme.

Ponzi scheme detection is to collect advertisements claiming high returns in various communities and forums, and capture their Bitcoin addresses for manual analysis of transactions to identify them~\cite{vasek2015there}. Later, Bartoletti et al.~\cite{bartoletti2020dissecting} proposed to calculate the number of character edits that convert one byte-code to another byte-code based on the Normalized Levenshtein distance~(NLD) as the contract similarity to identify Ponzi schemes. However, the amount of smart contract code limits the above methods, which need to rely on known codes to identify similar codes. And then, Chen et al.~\cite{chen2018detecting, chen2019exploiting} proposed a method based on machine learning to identify schemes, mainly analyzing the characteristics of contract transactions and counting contract byte codes. Furthermore, Fan et al.~\cite{fan2020expose} improved the method of combining feature engineering and machine learning, and used the idea of ordered enhancement to train the Ponzi detection model. Therefore, the ordered target statistics method can directly process category features and avoid the prediction offset caused by target leakage. The above identification of Ponzi schemes is mostly based on feature engineering, e.g., manually extracts the transaction features of the contract, counts the feature frequency of the contract byte-code, and combines the acquired features. In this paper, we propose a Ponzi scheme detection method that combined with manual contracts' features and topological structure of the transaction network by GCN, which can effectively detect Ponzi schemes.

\subsection{Network Embedding Methods}
Network embedding methods~\cite{cai2018comprehensive} have received much attention over the past decades, which convert each node into a low-dimensional vector and make each vector retain as much of the original structural information and topological structure as possible. There are many network embedding methods, which are mainly classified into the following categories, i.e., matrix factorization, deep learning, and other miscellaneous strategies.

Matrix factorization is based on global paired statistical information to learn the similarity of representations. However, matrix factorization methods have high time and space complexity, which is unsuitable for large networks. Deep learning~(DL) can automatically recognize useful representations of complex network structures. Generally, DL can be divided into two categories, i.e., deep learning based on random walk and deep learning without random walk. DeepWalk~\cite{perozzi2014deepwalk} and node2vec~\cite{grover2016node2vec} are two classic examples using random walks on networks to learn node representation. Deepwalk is a random walk embedding algorithm that combines the Skip-Gram model to obtain node representations. While node2vec is an improved version of DeepWalk, which adjusts the super parameters of $p$, $q$, to make embedding balanced in structural equivalence and homophily. GCN~\cite{kipf2017semi} is a classic deep learning method that generates node representation by aggregating its own features and neighbors' features and stacking multiple graph convolutional layers to extract high-level node representations. Differ to the above methods, LINE~\cite{tang2015line} preserves both the local and global network structures modeling node co-occurrence probability and conditional probability. Considering the high expressiveness and learning ability of GCN, we propose a GCN-based model to well aggregate with the nodes features and topological structure of transaction networks to make Ponzi scheme detection.

\section{Methodology} \label{sec:Methodology}
\begin{figure}[!t]
\centering
\includegraphics[width=\linewidth]{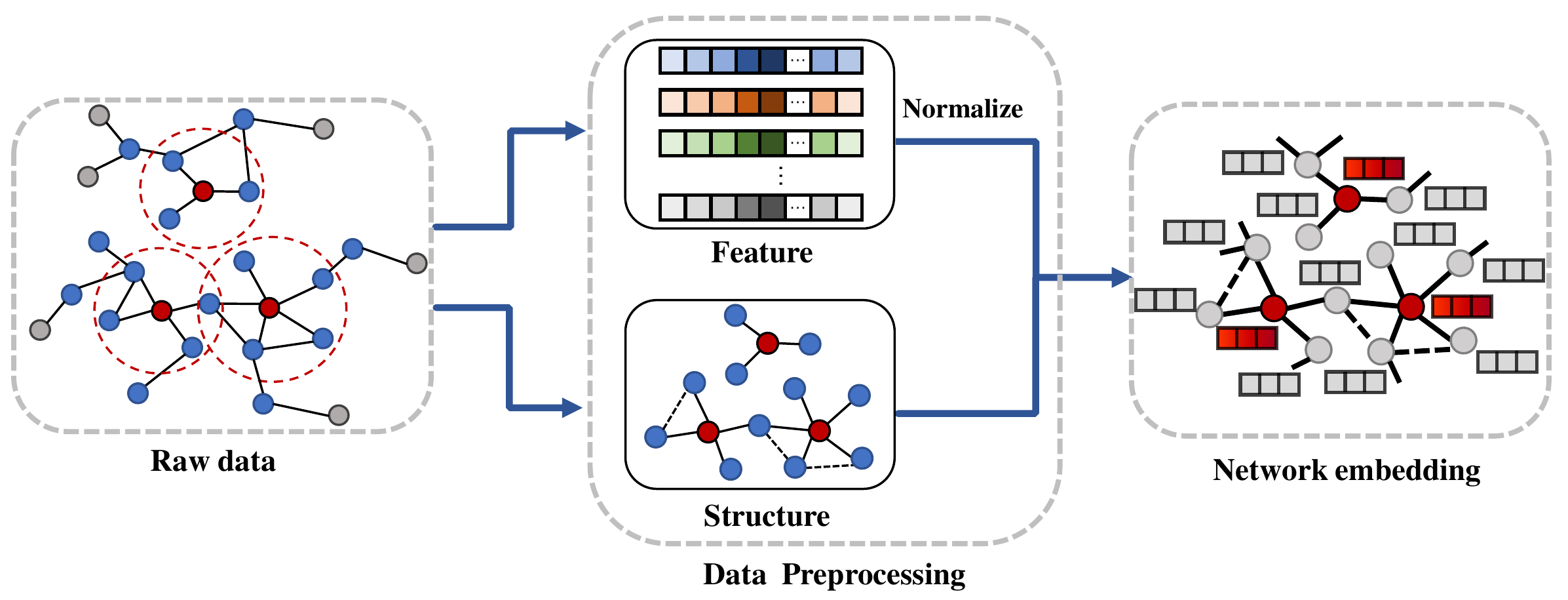}
\caption{Overall framework of our proposed model.}
\label{fig:Frame}
\end{figure}

In this section, we first give a detailed description of data collection and data preprocessing, and then extract the following meta-features for the smart contracts, which can form more complex features. We also represent the details of our proposed model for Ponzi scheme detection. The framework of our proposed model is illustrated in Fig.~\ref{fig:Frame}.

\subsection{Data Collection \& Preprocessing}\label{sec:DataCollection}
Data collection is the basis of transaction network analysis. Thanks to the openness of Ethereum, we can easily access transaction records and smart contracts. In this work, we first collect 50 labeled Ponzi schemes from an authoritative website,~\textit{Ethercan}\footnote{Etherscan:~\url{etherscan.io}}, which reports various illegal behaviors on Ethereum. However, only 50 contracts are labeled as Ponzi schemes, making it difficult to find the common characteristics of these smart contracts. We further explore more Ponzi schemes in \textit{Xblock}\footnote{Xblock:~\url{http://xblock.pro/ethereum/}}, which collects datasets for researchers to autonomously access Ethereum meta-datasets.

\begin{figure}[!t]
\centering
\includegraphics[scale=0.4]{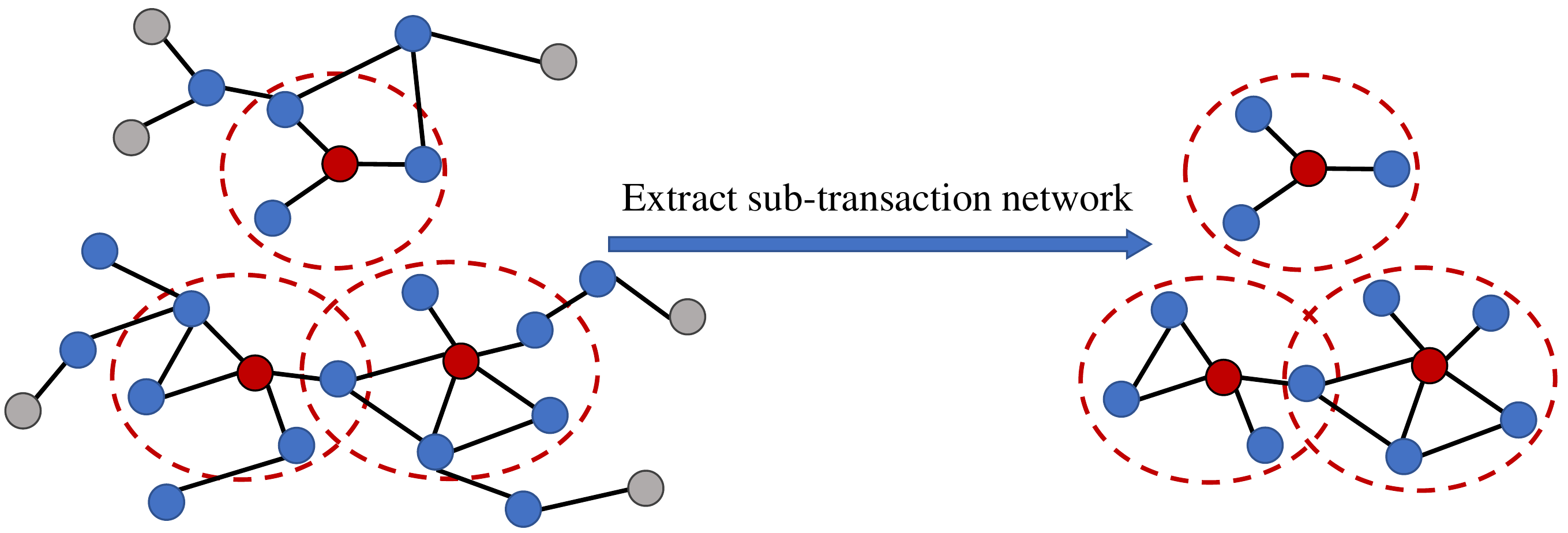}
\caption{Schematic illustration of the sub-transaction network.}
\label{fig:k_neborhood}
\end{figure}

Even though many fraudulent activities are very active on Ethereum, the fraudulent activities of Ponzi contracts still account for a small part, which shows that the number of Ponzi contracts and normal contracts are extremely imbalanced. Therefore, we take the number of Ponzi contracts as the guideline, selecting an equal number of normal contracts in all contract transactions, and then forming datasets for the following experiments. We collect transaction records from March 2015 to March 2020. It is worth noting that the transaction records are extremely large, and thus we ascertain a number of target smart contracts~(Ponzi contracts and normal contracts) and then obtain their transactions from all Ethereum transaction records to make subgraphs for subsequent experiments. As shown in Fig.~\ref{fig:k_neborhood}, we randomly sample centered contracts to obtain their 1st-order neighbors and the transaction records between all of them. In other words, we extract the transaction network into a sub-transaction network~(the extracted subgraph is highly correlated with these target contracts), and then analyze and identify Ponzi schemes from a network perspective.

For simplicity, the extracted sub-transaction network can be defined as $G=(V,E)$, where $V=\{v_{1},\cdots,v_{n}\}$ refers to a set of nodes, and $(v_i,v_j) \in E$ represents the links among the nodes $v_{i}$ and $v_{j}$. Considering the input of our deep model is a matrix, we denote the adjacency matrix as $A \in \mathbb{R}^{n \times n}$. The corresponding degree matrix and feature matrix refer to $D_{i i}=\sum_{j} A_{i j}$ and  $X \in \mathbb{R}^{n \times d}$, respectively, where $n$ is the number of nodes, $d$ is the dimension of the feature vector, and each row $x_{i} \in \mathbb{R}^d$ is the feature vector of node $v_{i}$.

\subsection{Contract Features}\label{sec:AccountFeatures}
Due to the fraudulent essence, Ponzi contracts have distinct characteristics compared with normal contracts. The inherent characteristics of a smart Ponzi scheme determine its special behavior, which can be used to determine whether it is a Ponzi smart contract or not. Since the main behavior of contracts can be reflected through transactions with other accounts, we manually investigate the transaction history of the target smart contracts. After obtaining all the transaction records, we can directly obtain two types of information: transaction amount and transaction time. In order to distinguish the nature of the transactions, we use the transaction direction to measure the transaction-in and transaction-out. For each smart contract, we combine the direction of the transactions to count the in-degree and out-degree of the target contract, i.e., the number of contract in and out transactions. For the transaction amount, we calculate the total, maximum, minimum, average, and variance of the target contract's transaction amount from the inflow and outflow of contract transactions. For transaction time, we specifically count the contracts' lifetime and mainly calculate the time interval between the initial contract creation time and the latest contract run time. In summary, we obtain 14 features, which can be used as the basic features of our proposed model.

\subsection{GCN}
We first introduce the GCN model and then illustrate how to use it for Ponzi scheme detection. GCN is a semi-supervised convolutional neural network~\cite{kipf2017semi} that can work directly on the network and take advantage of their structural information. Graph convolution has a wide range of applicability, which is suitable for nodes and graphs of any topology. It can simultaneously learn node feature information and structure information end-to-end. After establishing sub-transaction networks and extracting contracts' features, we use a 3-layer GCN model with softmax function to learn node representations for Ponzi scheme detection as follows:
\begin{equation}
    Z=\operatorname{softmax}\left(\hat{A}\operatorname{ReLU}\left(\hat{A}\operatorname{ReLU}\left( \hat{A}XW^{\left(0\right)} \right) W^{\left(1\right)} \right) W^{\left(2\right)} \right)     
\end{equation}
where $Z \in \mathbb{R}^{n \times y}$ is the probability distribution of classification, $n$ is the number of network nodes, and $y$ is the dimension of node labels.

\section{Experiments} \label{sec:Experiment}
In this section, we present our experimental results. We first introduce the datasets, followed by experimental settings and evaluation metrics. After that, we show the experimental results with discussion. Finally, we analyze the importance of features. 

\subsection{Dataset}
We collect 250 labeled Ponzi smart contracts which are the target of our detection model, and randomly select 250 unlabeled smart contracts as the outliers. With these labeled and unlabeled smart contracts being the central nodes, we extract their 1st-order neighbors and the transaction records between all of them through the API of~\textit{Etherscan} and then splice them into a sub-transaction network. In our experiments, we repeat the random selection procedure of labeled and unlabeled nodes three times and thus obtain three sub-transaction networks. The basic statistics of these datasets are presented in Table~\ref{tab:dataset}.
\begin{table}[!t]
\setlength{\tabcolsep}{5mm}
\centering
\renewcommand{\arraystretch}{1.2}
\setlength{\abovecaptionskip}{0pt}
\setlength{\belowcaptionskip}{5pt}
\caption{Basic topological features of the sub-transaction networks. $|V|$ and $|E|$ are the numbers of nodes and edges, respectively, $K$ is the average degree, and $C$ is the clustering coefficient.}
\begin{tabular}{ccccc}
\toprule
        Dataset & $|V|$   & $|E|$     & $K$       & $C$                        \\ \midrule
        DS1      & 34699   & 99745     & 7.562    & 0.386                      \\
        DS2      & 17980   & 64299     & 6.914    & 0.383                      \\
        DS3      & 23065   & 92200     & 7.222    & 0.418                      \\ \bottomrule
\end{tabular}
\label{tab:dataset}
\end{table}

\subsection{Baselines and Experimental Setup}
To validate the effectiveness of our proposed model, we compare it with LINE, as a widely used baseline network embedding method, as well as two popular network embedding approaches based on the random walk, i.e., DeepWalk and node2vec. Furthermore, we compare the results of the non-embedding methods which don't consider the structural information, but only the extracted features. In particular, the baselines are introduced as follows.
\begin{itemize}
\item[$\bullet$] \textbf{LINE} is suitable for large-scale networks, which preserves both the local and global network structures to model node co-occurrence probability and conditional probability, respectively. It defines the concept of 1st-order and 2nd-order similarity and enhances the expressive ability of embedding.
\item[$\bullet$] \textbf{DeepWalk} is a proximity-based embedding method that obtains node context via random walk, which uses Skip-Gram model and uniform random walks to learn the neighborhood structure of the graph.
\item[$\bullet$] \textbf{node2vec} further exploits a flexible neighborhood sampling strategy, i.e.,  Breadth-first Sampling~(BFS) and Depth-first Sampling~(DFS). It has two hyper-parameters to choose a proper balance between BFS and DFS to preserve community structure based on homophily as well as a structural equivalence between nodes.
\item[$\bullet$] \textbf{Feature} only consider the extracted features as illustrated in Section~\ref{sec:AccountFeatures}.
\end{itemize}

To ensure a fair comparison, we implement DeepWalk and node2vec using OpenNE\footnote{OpenNE:~\url{github.com/thunlp/openne}}~(an open-source package for network embedding). For all embedding methods, we follow the literature by setting the dimension $d=128$. For random walk embedding methods, we set hyper-parameters as follows: the size of window $k=10$, the length of walk $l=80$, and walks per node $r=10$. For node2vec, we grid search over $p, q \in \{0.5, 1, 2\}$. For LINE, we use 2nd-order-proximity and set other parameters to the provided defaults. As for our proposed model, we closely follow the framework of Kipf\footnote{~\url{https://github.com/tkipf/gcn}} with 3-layer GCN, the dimensionality of output is fixed at 32, set the maximum number of epochs to be 1000, and dropout is set 0.5. We perform 5-fold cross-validation across all methods and datasets with Random Forest. To evaluate the performance of different methods in terms of Ponzi scheme detection, we consider three evaluation metrics, namely, Precision, Recall, and F1-score. Experiments are repeated for 5 random seed initializations and the average performance is reported.

\subsection{Results and Discussions}
\textbf{Classifier Performance.} Classifier selection of Ponzi scheme detection is an important factor affecting detection performance. Therefore, we consider several widely considered classifiers as baselines, namely, Logistic Regression~(LR), Support Vector Machine~(SVM), ADAM optimizer(ADAM), and Random Forest~(RF). Here we compare the results of the non-embedding methods which only consider extracted features as illustrated in Section~\ref{sec:AccountFeatures}. The detection results of different classifiers are compared in Table~\ref{tab:feature classification}. We can observe that the Random Forest classifier~(RF) has the best performance among the above classifiers as it is more suitable for Ponzi scheme detection, and thus we select it as the classifier in this work.
\begin{table}[!t]
\caption{Performance of different classifiers using extracted features.}
\renewcommand{\arraystretch}{1.2}
\setlength{\abovecaptionskip}{0pt}
\setlength{\belowcaptionskip}{5pt}
\centering
\begin{tabular}{m{2.3cm}<{\centering}m{2.3cm}<{\centering}m{2.3cm}<{\centering}m{2.3cm}<{\centering}m{2.3cm}<{\centering}}
        \toprule
        Metric                     & Classifier & DS1             & DS2             & DS3             \\ \hline
        \multirow{4}{*}{Precision} & LR         & 0.6779          & 0.6777          & 0.6644          \\
                                   & SVM        & \textbf{0.8268} & 0.7784          & 0.8205          \\
                                   & ADAM        & 0.7812          & 0.8222          & 0.8420          \\
                                   & RF         & 0.8042          & \textbf{0.8364} & \textbf{0.8697} \\ \hline
        \multirow{4}{*}{Recall}    & LR         & 0.5389          & 0.5317          & 0.5408          \\
                                   & SVM        & 0.5890          & 0.5475          & 0.5947          \\
                                   & ADAM        & 0.7491          & 0.7088          & 0.7325          \\
                                   & RF         & \textbf{0.7409} & \textbf{0.7421} & \textbf{0.7777} \\ \hline
        \multirow{4}{*}{F1-score}  & LR         & 0.5982          & 0.5900          & 0.5872          \\
                                   & SVM        & 0.6753          & 0.6410          & 0.6858          \\
                                   & ADAM        & 0.7633          & 0.7594          & 0.7785          \\
                                   & RF         & \textbf{0.7700} & \textbf{0.7844} & \textbf{0.8183} \\ \bottomrule
\end{tabular}
\label{tab:feature classification}
\end{table}

\textbf{Performance Comparison.} We compare our proposed model with the four baseline methods on the performance metrics Precision, Recall, and F1-score. Table~\ref{tab:methods} shows the performance of all the compared methods on the Ponzi scheme task. It is particularly necessary to be able to detect potential risks in time and remind users, even if these risks may not be a Ponzi scheme after passing specific tests. Such risk reminders are common in daily life, so we may be more inclined to recall accounts suspected of the Ponzi scheme. Therefore, we are most concerned about recalls among all indicators in this work. The performance of the feature-only method is the worst across all datasets. Compared with the model that only uses manual features, the precision of LINE, Deepwalk and node2vec are all significantly improved. A reasonable explanation is that the addition of structural information can indeed learn more informative representations and thus boost the classification. Moreover, we can observe that considering the combination of features and GCN leads to better classification performance than using only the extracting features. In particular, our proposed model consistently yields the generally best performance on all datasets. This result indicates that GCN can perceive the nearby information and well integrates node features into the topological structure, which achieves decent classification performance. 

\begin{table}[!t]
\caption{Performance of different network embedding methods.}
\centering
\renewcommand{\arraystretch}{1.2}
\setlength{\abovecaptionskip}{0pt}
\setlength{\belowcaptionskip}{5pt}
\begin{tabular}{m{2.3cm}<{\centering}m{2.3cm}<{\centering}m{2.3cm}<{\centering}m{2.3cm}<{\centering}m{2.3cm}<{\centering}}
        \toprule
        Metric                     & Method      & DS1             & DS2             & DS3             \\ \hline
        \multirow{5}{*}{Precision} & Feature     & 0.8042          & 0.8364          & 0.8697          \\
                                   & LINE        & 0.9351          & 0.9015          & 0.8714          \\
                                   & Deepwalk    & 0.9397          & \textbf{0.9450} & 0.8729          \\
                                   & node2vec    & \textbf{0.9444} & 0.9351          & \textbf{0.9313} \\
                                   & GCN+Feature & 0.8686          & 0.8766          & 0.8564          \\ \hline
        \multirow{5}{*}{Recall}    & Feature     & 0.7409          & 0.7421          & 0.7777          \\
                                   & LINE        & 0.7307          & 0.5897          & 0.5823          \\
                                   & Deepwalk    & 0.5241          & 0.5235          & 0.5834          \\
                                   & node2vec    & 0.6961          & 0.6921          & 0.6699          \\
                                   & GCN+Feature & \textbf{0.9159} & \textbf{0.9159} & \textbf{0.9409} \\ \hline
        \multirow{5}{*}{F1-score}  & Feature     & 0.7700          & 0.7844          & 0.8183          \\
                                   & LINE        & 0.8111          & 0.7079          & 0.6874          \\
                                   & Deepwalk    & 0.6553          & 0.6634          & 0.6943          \\
                                   & node2vec    & 0.7986          & 0.7891          & 0.7780          \\
                                   & GCN+Feature & \textbf{0.8907} & \textbf{0.8940} & \textbf{0.8963} \\ \bottomrule
        \end{tabular}
\label{tab:methods}
\end{table}

\textbf{Feature Analysis.} To further understand the discriminative power of the extracted features, we investigate the feature importance based on the Random Forest mean decreasing impurity as shown in Fig.~\ref{fig:feature importance}. Next, we analyze why some of these features are important.
\begin{itemize}
\item[$\bullet$] \textit{count\_in\//count\_out} reflects the number of transactions between two accounts. These features are more important than others, and the reason is relatively easy to understand. For a Ponzi smart contract, a natural phenomenon is that the Ponzi scheme uses the funds of new investors to subsidize the old investors. However, with the exposure of the Ponzi scheme, the number of transactions will suddenly decrease or even disappear. 
\item[$\bullet$] \textit{values\_max\_in\//values\_max\_out} represents the largest transaction input~(output) amount in the current contract. Ponzi scheme attracts investors to transfer money to it, which can accumulate a lot of wealth. Meanwhile, the Ponzi scheme will quickly transfer a lot of wealth to another account to avoid scam collapse. Due to the high cost of Ether currency, the transaction amount of most normal contracts is small. Therefore, the maximum value of the transaction can be used to distinguish Ponzi smart contract precisely.
\item[$\bullet$] \textit{lifetime\_in\//lifetime\_out} represents the smart contracts' lifetime. The experimental results show that this feature can play a certain role but not the most critical feature. This is mainly because only a small number of smart contracts are frequently used and remain active, while most of them have a short lifetime in Ethereum. Ponzi scheme is often profitable in a short time, it will no longer be used by investors once reported or destroyed in time, resulting in a similar lifetime between normal contracts and Ponzi contracts.
\end{itemize}
These results reveal a noteworthy phenomenon that manually features can make different views of the smart contract from different angles. Therefore, our proposed model incorporates the extracted features that can facilitate Ponzi scheme detection.

\begin{figure}[!t]
\centering
\includegraphics[width=0.8\linewidth]{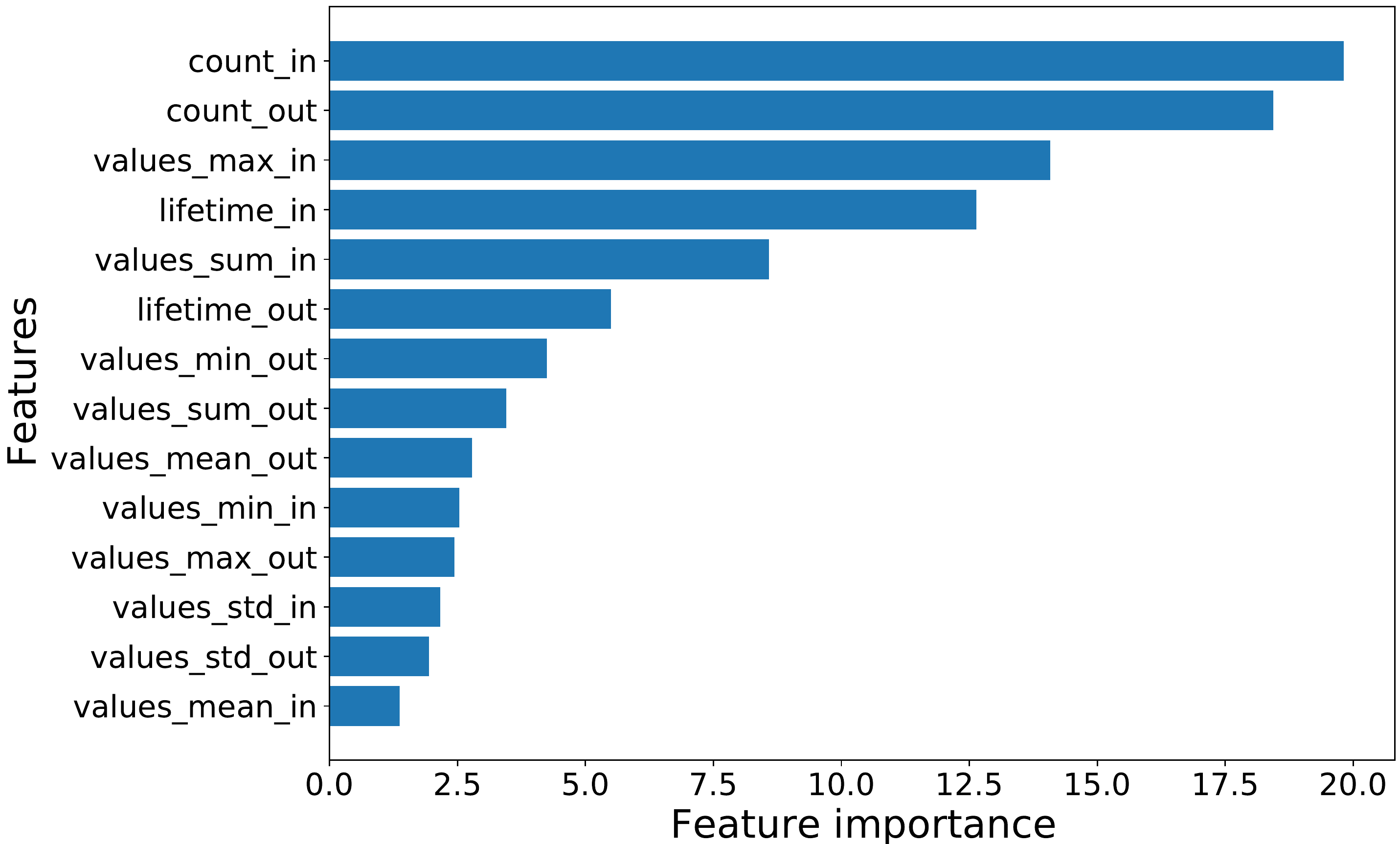}
\caption{The importance of the 14 extracted features.}
\label{fig:featureimportance}
\end{figure}

\section{Conclusion} \label{sec:Conclusion}
In Ethereum, various scams are rampant and the Ponzi scheme is the most serious threat to the financial security of users involved. To deal with this issue, in this paper, we collect transaction records from Ethereum and detect Ponzi schemes from a network perspective. We propose a GCN model that incorporates extracted features to detect and identify Ponzi smart contracts. Compared with general machine learning methods, the experimental results indicate that our proposed model performs better than a series of baselines to identify the Ponzi schemes. In the future, we plan to further extend our proposed model to detect Ponzi schemes in time. 

\subsubsection*{Acknowledgments.} Acknowledgements. The authors would like to thank all the members in the IVSN Research Group, Zhejiang University of Technology for the valuable discussions about the ideas and technical details presented in this paper. This work was partially supported by the National Key R\&D Program of China under Grant No. 2020YFB1006104, by the National Natural Science Foundation of China under Grant No. 61973273, by the Zhejiang Provincial Natural Science Foundation of China under Grant No. LR19F030001, by the Ministry of Public Security’s Research Project ``Research and Demonstration Application of Key Technologies of Criminal Social Network Model", and by the Special Scientific Research Fund of Basic Public Welfare Profession of Zhejiang Province under Grant LGF20F020016.

\bibliography{ref} 
\bibliographystyle{ieeetr}

\end{document}